\documentclass[9pt,twocolumn,twoside,nostamp,dvipsnames]{pnas-new}
\templatetype{pnasresearcharticle}

\usepackage[utf8]{inputenc}
\usepackage[T1]{fontenc}
\usepackage{microtype}
\usepackage{graphicx}
\usepackage{amsmath}
\usepackage{bm}
\usepackage{subcaption}
\usepackage{aas_macros}

\title{Multifield Cosmology with Artificial Intelligence}

\author[a,b,1]{Francisco Villaescusa-Navarro}
\author[c,b]{Daniel Angl\'es-Alc\'azar}
\author[b,d]{Shy Genel}
\author[b,a]{David N. Spergel}
\author[b]{Yin Li}
\author[e,b]{Benjamin Wandelt}
\author[a]{Andrina Nicola}
\author[f]{Leander Thiele}
\author[b,g]{Sultan Hassan}
\author[a]{Jose Manuel Zorrilla Matilla}
\author[h,i]{Desika Narayanan}
\author[j,g,k]{Romeel Dave}
\author[l]{Mark Vogelsberger}

\affil[a]{Department of Astrophysical Sciences, Princeton University, Peyton Hall, Princeton, NJ, 08544, USA}
\affil[b]{Center for Computational Astrophysics,
Flatiron Institute, 162 5th Avenue, New York, NY, 10010, USA}
\affil[c]{Department of Physics, University of Connecticut, 196 Auditorium Road, U-3046, Storrs, CT, 06269, USA}
\affil[d]{Columbia Astrophysics Laboratory, Columbia University, 550 West 120th Street, New York, NY, 10027, USA}
\affil[e]{Sorbonne Universite, CNRS, UMR 7095, Institut d’Astrophysique de Paris, 98 bis boulevard Arago, 75014 Paris, France}
\affil[f]{Department of Physics, Princeton University, Princeton, NJ 08544, USA}
\affil[g]{Department of Physics \& Astronomy, University of the Western Cape, Cape Town 7535, South Africa}
\affil[h]{Department of Astronomy, University of Florida, 211 Bryant Space Sciences Center, Gainesville, FL, USA}
\affil[i]{University of Florida Informatics Institute, 432 Newell Drive, CISE Bldg E251, Gainesville, FL, USA}
\affil[j]{Institute for Astronomy, University of Edinburgh, Royal Observatory, Edinburgh EH9 3HJ, UK}
\affil[k]{South African Astronomical Observatories, Observatory, Cape Town 7925, South Africa}
\affil[l]{Kavli Institute for Astrophysics and Space Research, Department of Physics, MIT, Cambridge, MA 02139, USA}

\leadauthor{Francisco Villaescusa-Navarro}

\correspondingauthor{\textsuperscript{1}
E-mail: villaescusa.francisco@gmail.com}

\authorcontributions{Author contributions:
FVN, DAA, SG, DNS, YL, and BW designed research;
FVN performed research;
FVN, DAA, SG, YL, DN, and RM contributed software;
FVN, DAA, SG, DNS, YL, BW, AN, LT, SH, JMZM, analyzed data;
FVN, DAA, SG, YL, BW, AN, LT, SH, JMZM, DN, and RM wrote the paper}

\authordeclaration{The authors declare no conflict of interest.}

\dates{This manuscript was compiled on \today}

\doi{\url{www.pnas.org/cgi/doi/10.1073/pnas.XXXXXXXXXX}}


\significancestatement{
A wealth of information from cosmological surveys remains out of reach owing to (1) the lack of an optimal method to extract that information and (2) the contamination introduced by uncertain astrophysical processes such as the impact of supernova explosions and growing supermassive black holes. This hinders our ability to learn about fundamental physics behind the laws and constituents of our Universe from cosmological data. We show for the first time that neural networks trained on thousands of galaxy formation simulations with varying cosmological and astrophysical models can help overcome these challenges by extracting the relevant cosmological information directly from multiple raw data fields while simultaneously marginalizing over uncertainties in astrophysical effects.
}

\begin{abstract}
Astrophysical processes such as feedback from supernovae and active galactic nuclei modify the properties and spatial distribution of dark matter, gas, and galaxies in a poorly understood way. This uncertainty is one of the main theoretical obstacles to extract information from cosmological surveys. We use 2,000 state-of-the-art hydrodynamic simulations from the CAMELS project spanning a wide variety of cosmological and astrophysical models and generate hundreds of thousands of 2-dimensional maps for 13 different fields: from dark matter to gas and stellar properties. We use these maps to train convolutional neural networks to extract the maximum amount of cosmological information while marginalizing over astrophysical effects at the field level. Although our maps only cover a small area of $(25~h^{-1}{\rm Mpc})^2$, and the different fields are contaminated by astrophysical effects in very different ways, our networks can infer the values of $\Omega_{\rm m}$ and $\sigma_8$ with a few percent level precision for most of the fields. We find that the marginalization performed by the network retains a wealth of cosmological information compared to a model trained on maps from gravity-only N-body simulations that are not contaminated by astrophysical effects. Finally, we train our networks on multifields  --  2D maps that contain several fields as different \textit{colors} or channels --  and find that not only they can infer the value of all parameters with higher accuracy than networks trained on individual fields, but they can constrain the value of $\Omega_{\rm m}$ with higher accuracy than the maps from the N-body simulations.
\end{abstract}

\begin{document}

\maketitle
\thispagestyle{firststyle}
\ifthenelse{\boolean{shortarticle}}{\ifthenelse{\boolean{singlecolumn}}{\abscontentformatted}{\abscontent}}{}

\dropcap{C}osmology is becoming a precise branch of astronomy. Quantities characterizing critical properties of our Universe, such as its age and geometry, are known with relatively high precision. On the other hand, some fundamental questions remain unanswered: What is the nature of dark energy? What are the neutrino masses? Is General Relativity correct on large scales? Data from upcoming cosmological surveys such as Euclid\footnote{https://www.euclid-ec.org}, eRosita\footnote{https://www.mpe.mpg.de/eROSITA}, DESI\footnote{https://www.desi.lbl.gov}, PFS\footnote{https://pfs.ipmu.jp}, Rubin observatory\footnote{https://www.lsst.org}, Roman observatory\footnote{https://roman.gsfc.nasa.gov}, SKA\footnote{https://www.skatelescope.org}, and Simons Observatory\footnote{https://simonsobservatory.org} are expected to shed light on these questions. Unfortunately, the procedure needed to extract the maximum amount of cosmological information from such surveys is currently unknown.

In the case of Gaussian density fields, such as the temperature anisotropies of the early Universe, the power spectrum is the summary statistic that completely characterizes the statistical properties of those fields. This means that, for these cases, the optimal procedure to extract the maximum amount of information is known: compute the power spectrum from the data and perform a maximum likelihood estimation of the parameters given the theory predictions.

This procedure is routinely used to perform parameter inference from observables such as the spatial distribution of galaxies on sufficiently large scales, since their distribution resembles a Gaussian density field. Unfortunately, even if the early Universe was a Gaussian density field, non-linear gravitational evolution would transform it into a non-Gaussian density field on small scales and at low redshift. Numerous studies have shown that extracting information using the power spectrum in this regime is suboptimal \citep[e.g.][]{Quijote, Samushia_2021, Gualdi_2021, Kuruvilla_2021, Bayer_2021,  Banerjee_2019, Changhoon_2019, Uhlemann_2020, Friedrich_2020, Massara_2020, Dai_2020, Allys_2020, Banerjee_2020, Banerjee_2021, Gualdi_2020, Gualdi_2021, Giri_2020, Bella_2020, Changhoon_2020, Valgiannis_2021, Bayer_2021}.

For generic non-Gaussian fields, the summary statistic(s) that fully characterizes their statistical properties is unknown. In other words, the procedure to extract the maximum information from these fields remains a mystery. This represents a major challenge that limits our ability to extract the maximum information from cosmological surveys. A second challenge arises because astrophysical processes, such as feedback from supernovae and active galactic nuclei (AGN), are expected to affect the distribution and properties of matter and galaxies in a poorly understood manner; for a generic statistic, the effect of baryonic physics is largely unknown. 

One possibility to tackle these problems is to use artificial neural networks to extract information from the field while simultaneously marginalizing over uncertain baryonic effects. We have recently demonstrated using Gaussian density fields that neural networks can achieve both goals \cite{Paco_2020b}. While many works have used neural networks to extract cosmological information from gravity-only N-body simulations \citep{Siamak_16, Schmelzle_17, Gupta_18, Ribli_19, Fluri_19, Ntampaka_19, Sultan_2019, Jose_2020, Niall_2020}, to our knowledge no such analysis has been performed using state-of-the-art cosmological hydrodynamic simulations with full galaxy formation physics and leveraging the information content of multiple baryonic fields simultaneously.

In this work, we make use of data from the Cosmology and Astrophysics with MachinE Learning Simulations (CAMELS) project \citep{CAMELS} to perform likelihood-free inference of cosmological parameters from 2D projected distributions of a variety of fields evolved in the simulations.
CAMELS contains more than 4,000 cosmological N-body and state-of-the-art hydrodynamic simulations, where each simulation represents a Universe with a different cosmological and astrophysical model. The hydrodynamic simulations have been run with two different codes and baryonic subgrid models: IllustrisTNG \citep{WeinbergerR_16a,PillepichA_16a} and SIMBA \citep{Dave2019_Simba} simulations, that include radiative cooling and heating, star formation, stellar evolution, galactic winds, the formation and growth of supermassive black holes, and feedback from AGN; this set of input physics can yield observationally-concordant galaxy populations in current models (Somerville+Dave 2015). The value of the cosmological and astrophysical parameters in the simulations is varied across a wide range, so that posterior location and width from simulated data from the middle of the range are not affected by priors.

From the hydrodynamic simulations, we generate tens of thousands of 2D maps for 13 different fields (from dark matter to gas and stellar properties), which are affected by astrophysical effects at very different levels. 
These maps represent a small area of $25\times25~(h^{-1}{\rm Mpc})^2$ and have a spatial resolution of $\simeq100~h^{-1}{\rm kpc}$. We then feed these maps to convolutional neural networks that are trained to perform likelihood-free inference on the value of the cosmological parameters while marginalizing over the astrophysical effects at the field level.

We show that neural networks not only can marginalize over baryonic effects at the field level, but they also achieve percent level constraints on the value of the cosmological parameters for most of the fields. This illustrates the enormous amount of cosmological information that is embedded in small scales in the fully non-linear regime. We compare our results to those obtained from maps that are not contaminated by astrophysical effects (the maps from the gravity-only N-body simulations), and we show that the network does not lose much information when performing the marginalization over the astrophysical effects. In addition, we train neural networks to perform likelihood-free inference on \textit{multifields}, defined as 2D maps that contain multiple fields simultaneously as different \textit{colors} or channels. We show for the first time that the multifield configuration provides the tightest constraints on most parameters despite the widely different astrophysical effects on each channel, outperforming the results from maps not contaminated by astrophysical effects. Finally, we discuss how our results can be used for forward models and the challenges associated with this method before they can be applied to data from cosmological surveys.

\begin{figure*}[!t]
\centering
\includegraphics[width=1.0\textwidth]{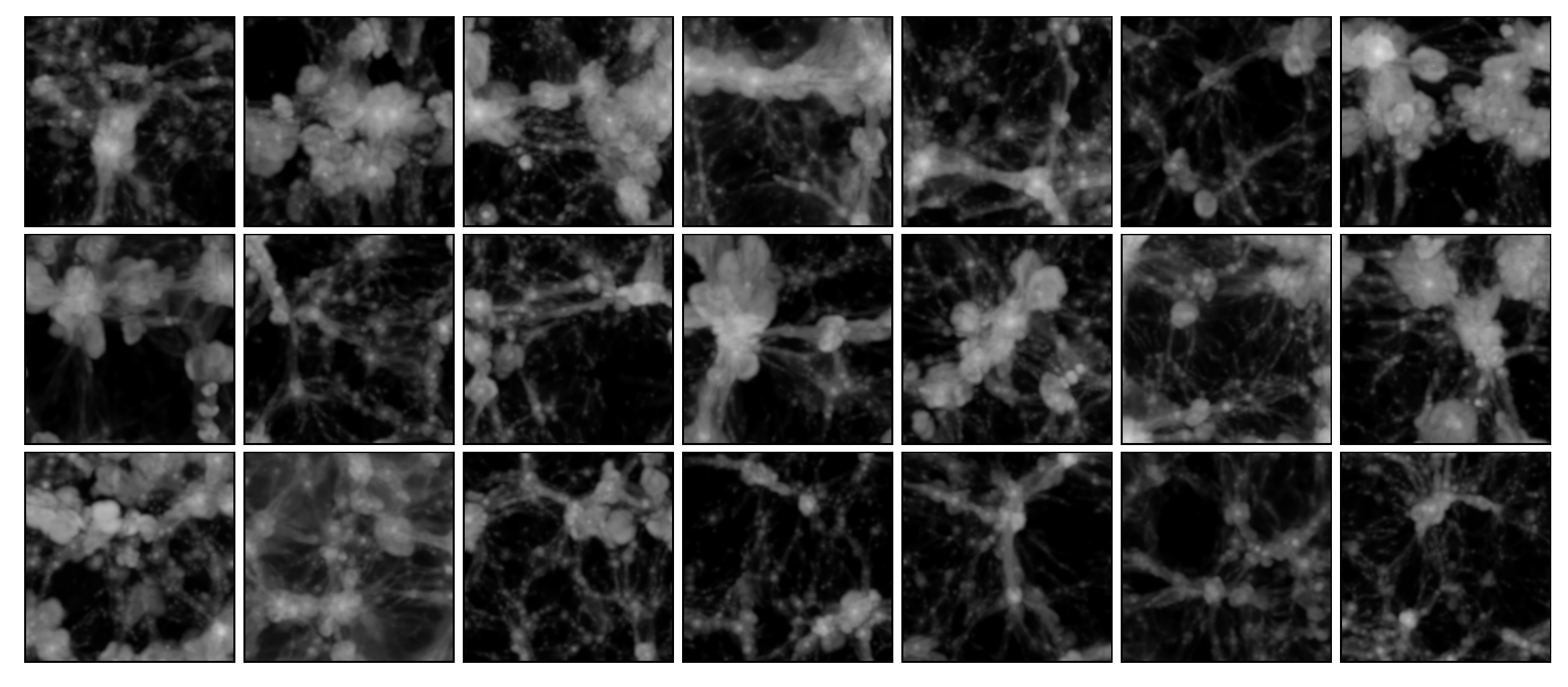}
\includegraphics[width=1.0\textwidth]{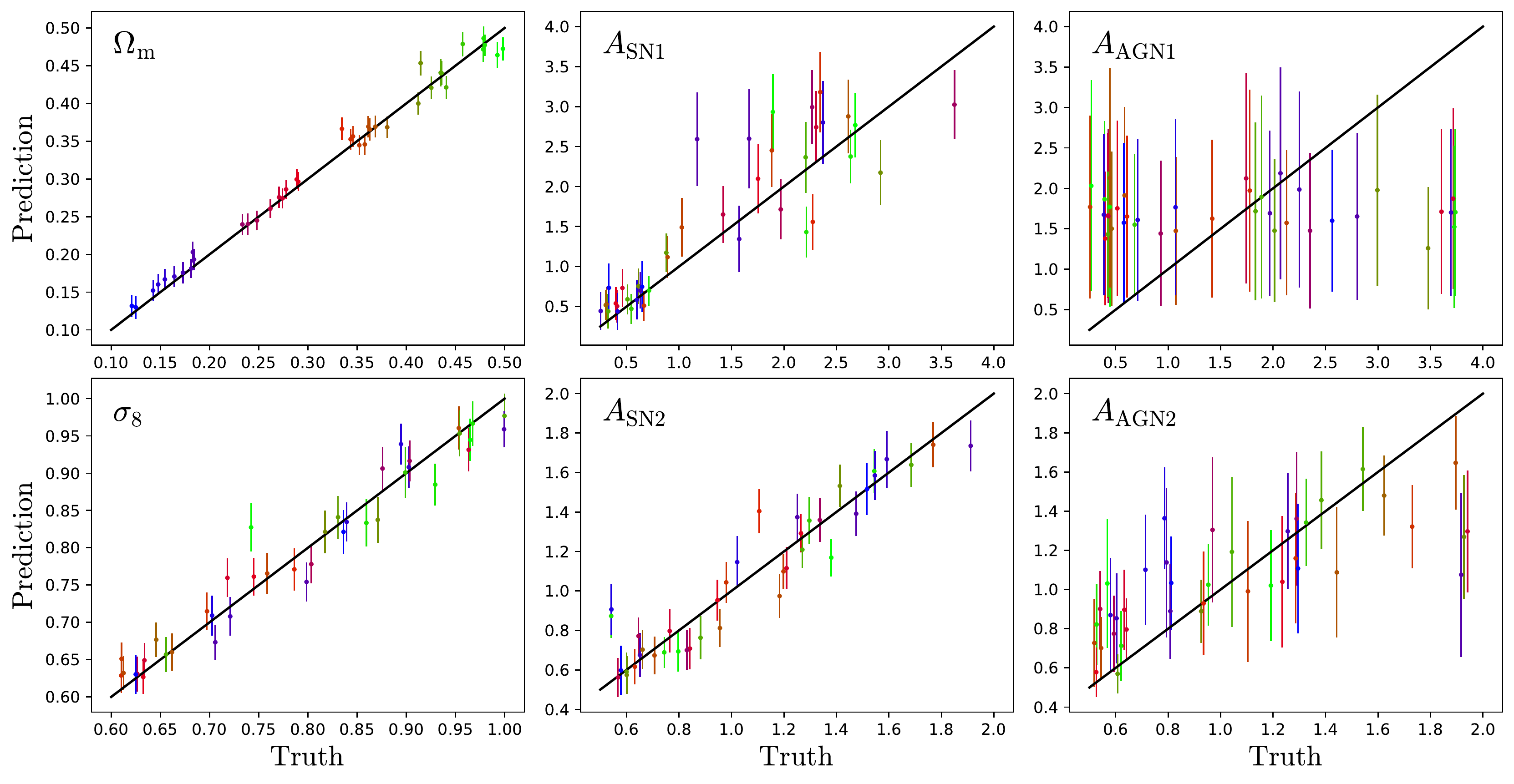}
\caption{We train a convolutional neural network to perform likelihood-free inference on the value of cosmological and astrophysical parameters from gas temperature maps. The top panels show examples of those maps. Each map contains $256\times256$ pixels over an area of $(25~h^{-1}{\rm Mpc})^2$ and has a different value of the cosmological and astrophysical parameters. With the network trained, we input to it 40 gas temperature maps from the test set (i.e. never seen by the network). For each of these maps the network outputs the mean and standard deviation of the  posterior. We show the results in the bottom panels; each panel shows the results for a different parameter. Points are color-coded according to their truth value of $\Omega_{\rm m}$. As can be seen, the network is able to infer the value of the cosmological parameters with high-accuracy while marginalizing over astrophysical effects at the field level.}
\label{fig:NN_pred}
\end{figure*}

\section*{Results}

We now present the main results of this work. We first illustrate in detail our method with one particular field: gas temperature. We then describe the results obtained for all other fields. Next, we present the results for the multifield maps and the N-body simulations. Unless otherwise stated, we only show results obtained from maps from the 1,000 IllustrisTNG hydrodynamic simulations belonging to the CAMELS project. Results are very similar if we instead use maps from the 1,000 SIMBA hydrodynamic simulations from CAMELS, which are run with a completely different code and utilize a very distinct subgrid physics implementation. We emphasize that unless otherwise stated, we show results when training networks on maps from the IllustrisTNG simulations and testing the model on maps from the same set of simulations\footnote{The maps in the testing set have not been seen by the neural network though.}. In the discussion section we describe the plausible outcome of training the networks on maps from one simulation suite and test them on maps from the other simulation suite.

\subsection*{Gas temperature} In the upper panels of Fig. \ref{fig:NN_pred} we show examples of one of the 13 fields considered in this paper: gas temperature. Each map represents a region of $25\times25~(h^{-1}{\rm Mpc})^2$ and contains $256\times256$ pixels; the value of the cosmological and astrophysical parameters is different for each map. In terms of the astrophysical parameters, $A_{\rm SN1}$ and $A_{\rm SN2}$ control the efficiency of galactic winds, while $A_{\rm AGN1}$ and $A_{\rm AGN2}$ characterize AGN feedback. Since each simulation was run with a different value of the random seed used to generate the initial conditions, each map also represents a different realization of the cosmic web. The regions where the gas is colder are shown in black, and they tend to correspond to cosmic voids. In white we show the regions where the gas is hotter, corresponding to places where dark matter halos hosting galaxies are located. These regions are typically surrounded by hot bubbles whose morphology depends on cosmology but also on the efficiency of the astrophysical processes. These maps clearly illustrate the diversity of cosmological and astrophysical models present in the CAMELS simulations.

Our goal in this paper is to infer cosmological and astrophysical parameters directly from these maps. To achieve this, we take 13,500 of these maps (training set) and train a neural network to predict the mean and standard deviation of the marginal posterior for each of the six cosmological and astrophysical parameters (see the Neural networks subsection below for details). Once the network is trained, we input to it 40 maps that it has never seen before (belonging to the test set) and show the results in the bottom panels of Fig. \ref{fig:NN_pred}. We find that the network is able to infer the value of the cosmological parameters with a surprisingly high accuracy: $\langle \delta \Omega_{\rm m}/\Omega_{\rm m}\rangle\simeq3.5\%$ and $\langle \delta \sigma_8/\sigma_8 \rangle\simeq5.3\%$. We emphasize that these constraints are obtained from the maps themselves (i.e. without using intermediate summary statistics) and incorporate the marginalization over astrophysical effects as well as the effects of cosmic variance.

For these maps, the network can also infer the value of the two parameters controlling supernova feedback, but with larger errorbars. For the AGN parameters, the network can only predict the prior mean for $A_{\rm AGN1}$, showing that those maps do not contain much information about that parameter, or alternatively, our network is not powerful enough to extract that information. Similarly, the network can only infer the value of $A_{\rm AGN2}$ with  large uncertainty.

\begin{figure*}[!t]
\centering
\includegraphics[width=1.0\textwidth]{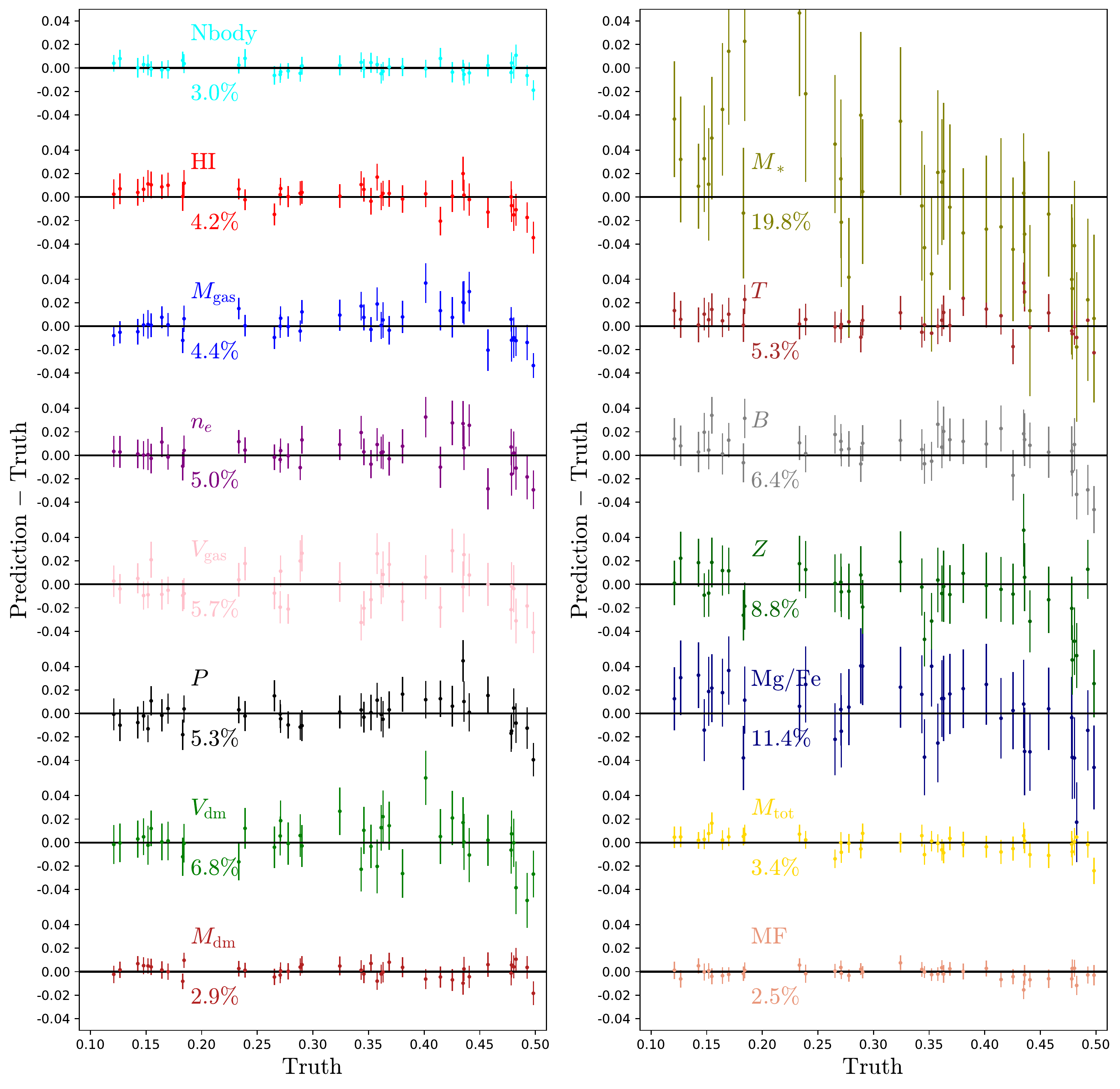}
\caption{We train neural networks to perform likelihood-free inference on $\Omega_{\rm m}$, $\sigma_8$, and the value of four astrophysical parameters from maps of 13 different fields and one multifield (see Fig. \ref{fig:images} for an illustration on the different fields). We use the trained networks to predict the mean and standard deviation of the posterior from a set of 40 maps of the test set of each field. In this plot we show the network prediction for $\Omega_{\rm m}$, minus its truth value, versus the true value, for all fields. As can be seen, the network is able to recover the true value in the different cases with different accuracy; while errors for total matter and dark matter are small, they are much larger for the stellar mass and the ${\rm Mg/Fe}$ field. However, even for the latter, the network is able to infer the value of $\Omega_{\rm m}$ with an error of $\langle \delta \Omega_{\rm m}/\Omega_{\rm m}\rangle $ of 11.4\% (see colored numbers). All results are from maps from hydrodynamic simulations, with the exception of N-body maps, that arise from gravity-only simulations. The ${\rm MF}$ maps represents images with 11 different \textit{colors} or channels: \textit{multifield}. Each channel contains a different property: $M_{\rm gas}$, $M_{\rm tot}$, $M_*$, $V_{\rm gas}$, ${\rm HI}$, $P$, $T$, $Z$, $n_e$, $B$, and ${\rm Mg/Fe}$. The derived constraints incorporate the effects of 1) marginalizing over astrophysical effects (with the exception of the N-body case) and 2) cosmic variance.}
\label{fig:LFI_Om}
\end{figure*}

\begin{figure*}
\centering
\includegraphics[width=0.98\textwidth]{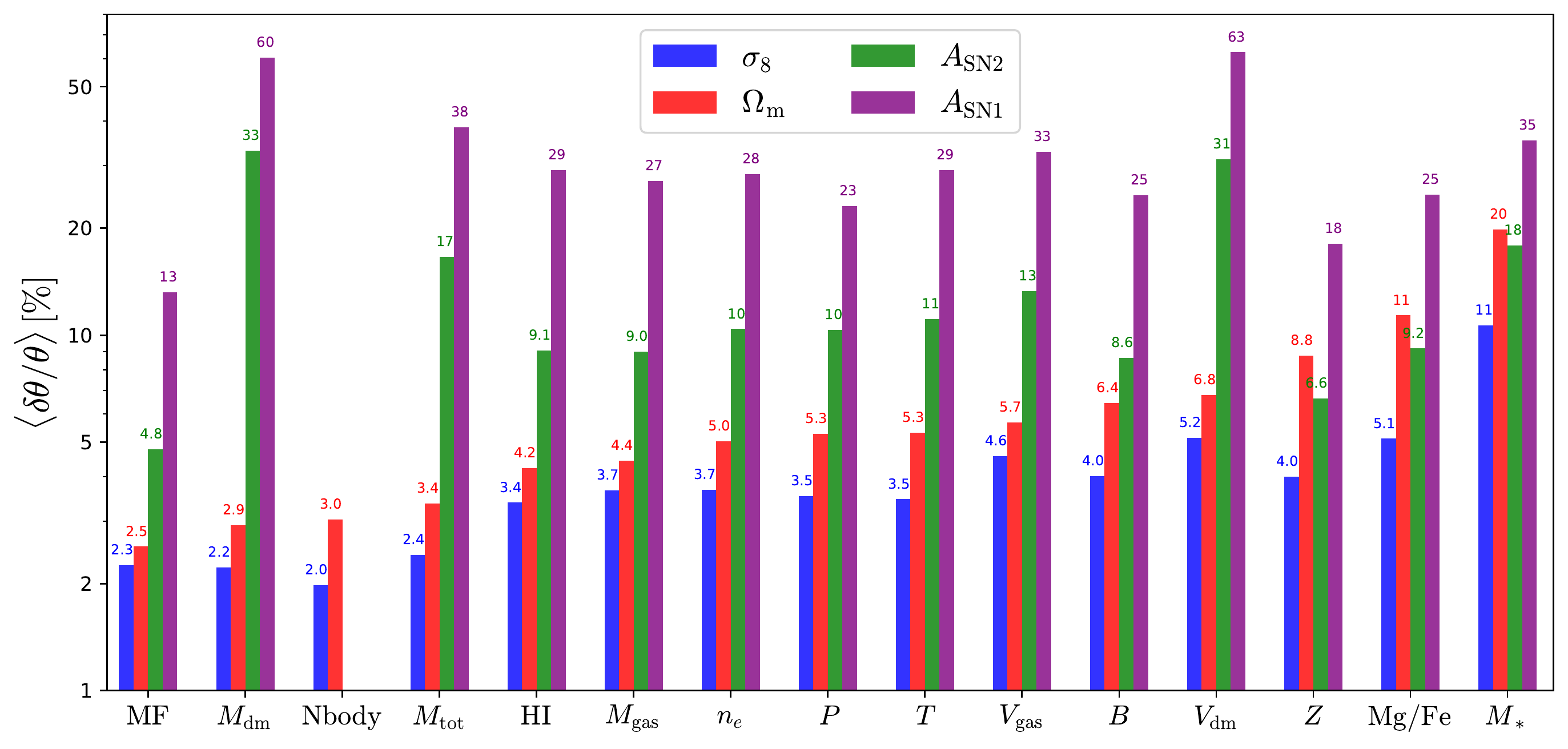}
\caption{We have trained convolutional neural networks to predict the mean ($\mu_\theta$) and standard deviation ($\sigma_\theta$) of the posterior for the parameters $\Omega_{\rm m}$, $\sigma_8$, $A_{\rm SN1}$, $A_{\rm SN2}$, $A_{\rm AGN1}$, and $A_{\rm AGN2}$ using as input different fields from the hydrodynamic simulations: $M_{\rm dm}$ (dark matter density), $M_{\rm tot}$ (total matter density), $n_e$ (electron number density), $T$ (gas temperature), $P$ (gas pressure), ${\rm HI}$ (neutral hydrogen density), $M_{\rm gas}$ (gas density), $V_{\rm dm}$ (dark matter velocity), $B$ (magnetic fields modulus), $V_{\rm gas}$ (gas velocity), $Z$ (gas metallicity), ${\rm Mg/Fe}$ (magnesium over iron ratio in the gas phase), and $M_*$ (stellar mass density), as well as the total matter from gravity-only N-body simulations (Nbody). We have also considered the case of a multifield that includes all fields above with the exception of $M_{\rm dm}$ and $V_{\rm dm}$, as different channels of the same input image; we denote these maps as ${\rm MF}$. This figure shows the mean relative accuracy, $\langle \sigma_\theta /\mu_\theta \rangle$ for all different fields and for four parameters; constraints on the AGN parameters are very poor in all cases and are not shown. Fields such as $M_{\rm dm}$ and $M_{\rm tot}$ contain a large amount of cosmological information while very little about astrophysics. On the other hand, fields such as $Z$ and ${\rm Mg/Fe}$ are good to constrain the value of the astrophysical parameters but not for the cosmological ones. In all cases, combining the fields as different channels of the same map yields tighter constraints than each individual field separately. Note that the constraints on $\Omega_{\rm m}$ from the multifield are better than those from the N-body maps, that are not affected by astrophysical effects.}
\label{fig:all}
\end{figure*}

\subsection*{All fields} We repeat the above exercise for each of the 13 different fields considered in this work, 1) gas density $M_{\rm gas}$, 2) gas velocity $V_{\rm gas}$, 3) gas temperature $T$, 4) gas metallicity $Z$, 5) gas pressure $P$, 6) electron number density $n_e$, 7) neutral hydrogen density ${\rm HI}$, 8) stellar mass density $M_*$, 9) dark matter density $M_{\rm dm}$, 10) dark matter velocity $V_{\rm dm}$, 11) magnetic fields $B$, 12) total matter density $M_{\rm tot}$, and 13) the ratio between the magnesium and iron mass content of gas ${\rm Mg/Fe}$. Furthermore, we also do the analysis on a particular multifield ${\rm MF}$ that contains images with 11 different \textit{colors}, or channels, (all fields above with the exception of dark matter density and dark matter velocity) and also on total matter density maps from the gravity-only simulations (${\rm Nbody}$), which, as opposed to all other fields, are not affected by astrophysical effects.

Fig. \ref{fig:LFI_Om} shows the results when  testing the different networks to infer the value of the parameter $\Omega_{\rm m}$. As can be seen, the network trained on some fields can infer the value of this cosmological parameter with higher accuracy than others. For instance, the dark matter density achieves an average relative accuracy, defined as $\langle \sigma_i/\mu_i\rangle$ where $\mu_i$ and $\sigma_i$ are the mean and standard deviation of the marginal posterior, of 2.9\% while for gas metallicity this number increases to 8.8\%. This is expected as the impact of astrophysical effects on the different fields is very different. We also note that we do not observe significant biases in the network predictions for all fields,  with the possible exception of the stellar mass density. 

Fig. \ref{fig:all} shows the average relative accuracy reached by the networks for all different fields and for the different parameters. We do not show the results for the AGN parameters as they are poorly constrained in all cases. Similarly to above, different fields are more sensitive to cosmology than others; the same is true for the astrophysical parameters.

We remind the reader that all the results are derived from present-day, or $z=0$, maps covering an area of $25\times25~(h^{-1}{\rm Mpc})^2$ and a resolution of $\simeq100~h^{-1}{\rm kpc}$. The constraints derived for the value of the cosmological parameters incorporate the effects of cosmic variance and the degradation due to the marginalization over astrophysical effects. 

\subsection*{Hydrodynamic fields} Of all the fields from the hydrodynamic simulations, $M_{\rm dm}$ and $M_{\rm tot}$ allow to place the tightest constraints on the value of $\Omega_{\rm m}$ and $\sigma_8$. We believe that this is because they are the ones least affected by astrophysical effects, and therefore the cosmological signal is stronger and less contaminated. Besides, these parameters are defined over the total matter density field, and the dark matter density is a good proxy for the total mass. For the fields related to gas properties, the neutral hydrogen density (${\rm HI}$), gas density ($M_{\rm gas}$) and electron density ($n_e$) are the most sensitive to cosmology, while the ratio between the magnesium and iron (${\rm Mg/Fe}$) is the least. The stellar mass density ($M_*$) yields the overall worst constraints. This could be due to the inherent sparsity associated with this field (most of the pixels are zero) and we believe that other kinds of networks, e.g. graph neural networks and deep sets, may perform better on those fields. 

Regarding the astrophysical parameters, we find that the gas metallicity yields the highest accuracy on both $A_{\rm SN1}$ and $A_{\rm SN2}$, followed by gas pressure, magnetic fields, and ${\rm Mg/Fe}$. One may expect that both the gas metallicity and the ratio of  magnesium over iron will exhibit a strong response to changes in the efficiency of feedback, much more than, e.g., the dark matter field, that can only be affected by these processes as a backreaction. This explains why the least sensitive fields to the astrophysical parameters are thus $M_{\rm dm}$, $M_{\rm tot}$, and $V_{\rm dm}$. It is however interesting to see that stellar mass density does not seem very sensitive to the astrophysical parameters either, although this could be due to the sparsity of the field. 

\subsection*{Multifield} In Figs. \ref{fig:LFI_Om} and \ref{fig:all} we see that the network trained on the maps that contain eleven channels (${\rm MF}$) can infer the value of $\Omega_{\rm m}$, $A_{\rm SN1}$, and $A_{\rm SN2}$ with higher accuracy than all other individual fields from the hydrodynamic simulations. This seems reasonable since the combination of these maps should contain more information than each individual field. We believe that in this case, the network may not only be using information from each individual field, but also making use of their cross-correlations, which not only are expected to be robust to systematic effects, but they can also break parameter degeneracies present in individual fields. The network can infer the value of $A_{\rm SN1}$ and $A_{\rm SN2}$ with $\simeq40\%$ higher accuracy than the most sensitive individual field, the gas metallicity. For $\Omega_{\rm m}$, the MF network accuracy is $\simeq35\%$ and $\simeq15\%$ better than the one achieved by the network trained on the $M_{\rm tot}$ and $M_{\rm dm}$ fields. On the other hand, for $\sigma_8$, the MF field yields a slightly worse accuracy than $M_{\rm dm}$ does, but better than $M_{\rm tot}$, though we believe that these differences are not statistically significant. We remind the reader that the MF maps do not contain explicit information about $M_{\rm dm}$ and $V_{\rm dm}$.

\subsection*{N-body fields} We find that the network trained on maps from the gravity-only simulations (N-body) achieves the highest accuracy on $\sigma_8$ ($\simeq2\%$) when compared with the fields from the hydrodynamic simulations. This result is expected as those maps are not contaminated by astrophysical processes, and therefore the network does not need to marginalize over them. On the other hand, the accuracy reached on $\Omega_{\rm m}$ is $3\%$, sightly worse than the one obtained on this parameter when training a network using the dark matter field of the hydrodynamic simulations ($2.9\%$). However, we do not believe that this difference is statistically significant.

What is more interesting is that for $\Omega_{\rm m}$ the network trained on the multifield maps achieves higher accuracy than the one trained on N-body maps. This may not seem physical, as the maps from the N-body simulations are not contaminated by astrophysical effects, but we believe several factors could explain these results. First, $\Omega_{\rm m}$ is the sum of $\Omega_{\rm dm}$ plus $\Omega_{\rm b}$, and hydrodynamic fields (e.g. gas properties) may carry more information about $\Omega_{\rm b}$ than their N-body counterparts. Second, the network may be exploiting the use of cross-correlations to beat down the errors from cosmic variance. In other words, with two fields sampling the same region, one can extract information from each individual field but also from their cross-correlation and cross-correlation coefficient. Third, the different channels (fields) may keep information about the evolutionary history in different ways that when used together may allow a better reconstruction than each individual field. These conditions cannot be satisfied by the maps from the N-body simulations, and may be the origin of the additional information the multifield is able to extract.

\section*{Discussion}
\label{sec:discussion}

In this section we first summarize the main conclusions of this work, discussing the implications of our results for the multifield maps. We then present the limitations of this methodology and discuss the road map needed to use this approach with real data.

The main conclusions of this work can be summarized as follows:
\begin{itemize}
\item Neural networks, trained to extract the maximum information while marginalizing over astrophysics effects at the field level, can infer the value of $\Omega_{\rm m}$ and $\sigma_8$ with a few percent precision from 2D maps covering a tiny area of $(25~h^{-1}{\rm Mpc})^2$. This result holds for 13 different fields from state-of-the-art hydrodynamic simulations that are affected by astrophysical effects in very different ways.
\item{The fields that allow a tighter constrain on the cosmological parameters are the dark matter density and total matter density, followed by neutral hydrogen density, gas density, and electron number density. The worst constraints are derived from the stellar mass density.}
\item{Unorthodox fields such as the gas-phase ratio between magnesium and iron ${\rm Mg/Fe}$ allow to constrain the value of $\Omega_{\rm m}$ and $\sigma_8$ with relatively high precision (11\% and 5.1\%, respectively) from the small 2D maps.}
\item{The fields that carry more information about the astrophysical model are the gas metallicity, the gas pressure, and the magnetic fields. The dark matter density and velocity fields yield instead the worst constraints on the astrophysical parameters.}
\item{Combining several fields, as different \textit{colors} of the same image -- \textit{multifield} -- allows one to retrieve the largest amount of cosmological and astrophysical information. We find that these fields contain more information than that accessible via maps that are not contaminated by astrophysical effects, namely the maps from the gravity-only simulations.}
\item{When comparing the results obtained by training networks on the total matter density from hydrodynamic and gravity-only simulations, we find that the N-body maps allow to place tighter constraints, but those from the hydrodynamic simulations are not much worse. This suggests that the network is extracting information in a complicated manner making use of small-scale features.}
\end{itemize}

It is interesting to note that while the multifield approach yields very significant improvements on the constraints of the parameters, they may be more modest than one could have expected. This may be due to the fact that all our fields are strongly correlated among themselves (see for instance Fig. \ref{fig:images}). While this is not good news from the point of view of obtaining tighter constraints to improve our understanding of fundamental physics, we can turn this around and explore it as an advantage to internally validate the method. For instance, one can use the subset of fields that are more sensitive to cosmology and astrophysics and determine the parameters with as high accuracy as possible from them. With those values, one can run simulations and take the fields not used to determine the parameters as predictions of the model, that can be directly contrasted with data. This approach will provide an internal cross-check to quantify the validity of the method.

\begin{figure}[t]
\centering
\includegraphics[width=0.49\textwidth]{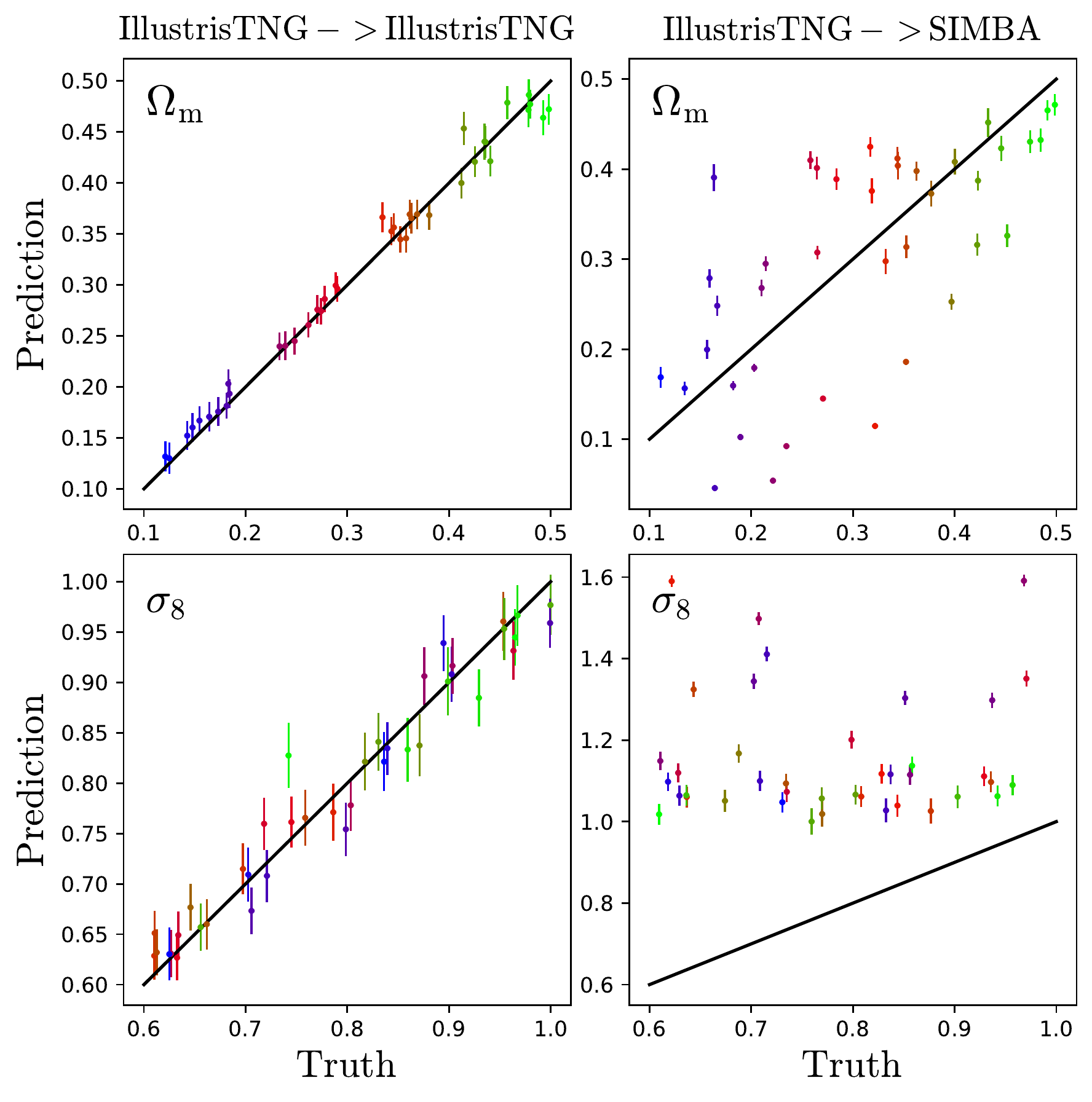}
\caption{We train a neural network on gas temperature maps from the IllustrisTNG simulations. We first test its accuracy using maps from the IllustrisTNG test set; results are shown in the left column. As can be seen, the network is able to constrain the value of $\Omega_{\rm m}$ and $\sigma_8$ with high accuracy. We also test the network using gas temperature maps from the SIMBA simulations, and show the results in the right column. In this case, the network fails dramatically when inferring the cosmological parameters. This behavior could be due to intrinsic differences between the simulations (see Fig. \ref{fig:differences}) or perhaps because the network is learning  unique features of each simulation.}
\label{fig:systematics}
\end{figure}

\subsection*{Limitations} One critical aspect to consider is the \textit{robustness} of this method. In other words, how does this method behave under the presence of systematic effects? To check this, we take the network that has been trained on gas temperature maps of the IllustrisTNG simulations and test it on gas temperature maps from the SIMBA simulations. We remind the reader that those two simulation suites are part of the CAMELS project, and differ in the method used to solve the hydrodynamic equations as well as the subgrid galaxy formation implementation. We show the results in Fig. \ref{fig:systematics}.

\begin{figure}[t]
\centering
\includegraphics[width=0.49\textwidth]{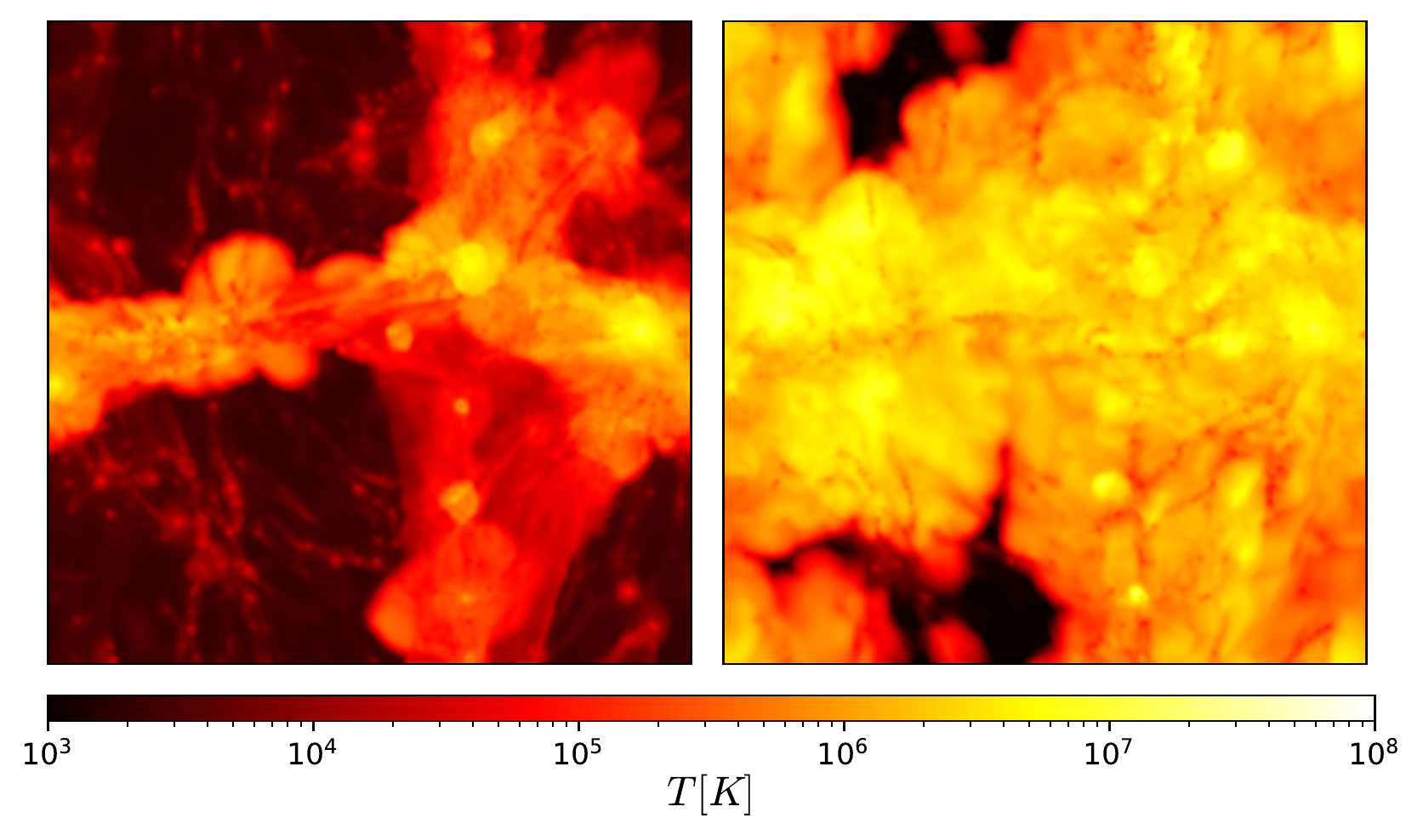}
\caption{We run two simulations with the same random seed and value of the cosmological and astrophysical parameters. This figure shows the projected gas temperature field in a slice of $25\times25\times5~(h^{-1}{\rm Mpc})^3$ when running the simulations using the IllustrisTNG (left) and SIMBA (right) models. As can be seen, differences are striking and arise due to the very different subgrid models used in these two simulations.}
\label{fig:differences}
\end{figure}

We find that under this setup, the network is not able to infer the correct value of the cosmological parameters; not only the mean of the posterior is off, but also the standard deviation of the posterior is incorrectly predicted. Note that we do not attempt to test the network on the value of the astrophysical parameters, as those are different among the IllustrisTNG and SIMBA simulations \cite[see][]{CAMELS}, so we do not expect them to work.

This behaviour can be due to several factors. First, it can be attributed to the large, intrinsic, differences between the IllustrisTNG and SIMBA simulations. We illustrate this point in Fig. \ref{fig:differences}, where we show gas temperature maps from two simulations with the same initial conditions and the same value of the cosmological and astrophysical parameters (fiducial model). As can be seen, the large-scale gas distribution in the SIMBA simulation is significantly hotter than that in IllustrisTNG, owing to the long range effect of collimated outflows driven by AGN \citep{Angles-Alcazar2017_BHfeedback,Dave2019_Simba}. These differences are a manifestation of the distinct subgrid models implemented in each simulation. 

On the other hand, one could have expected that given the large range of variation in the value of the astrophysical parameters, the two simulations should overlap over a relatively large volume in parameter space that would in any case be marginalized over. However, this may not be the case; the range of variation in the parameters may not be wide enough, or the parameters that are kept fixed may be producing differences large enough to prevent the overlap of the two simulation sets. Another possibility is that the network is learning unique features of each simulation (e.g. numerical artifacts) and therefore it fails when tested on a different simulation than the one used to train the network. Exploring these different hypotheses is beyond the scope of this work and we plan to address this in future work.

While the network trained on gas temperature maps is not robust to systematic effects, this conclusion is not generic to all other fields. In a companion paper \citep{Robust_marginalization}, we show that for some fields, specifically the total matter density field, the neural networks do not suffer from this problem, and training them on one suite of simulations does allow them to infer the true cosmology from a completely different simulation suite. To address the robustness issued raised above, it will be beneficial to explore in future work the use of contrast learning \cite{Contrast_learning} to force the network to learn robust summary statistics of the data.

Our results are promising and open the door to extracting information from cosmological surveys targeting different fields without the need to use summary statistics or apply a cut in scale to avoid contamination from astrophysical effects. Moreover, our results with multifields show that a large fraction of information can in principle be retrieved from a subset of all fields. This may allow us to use some fields to infer the cosmological and astrophysical model while using other fields for cross-checking and validating the entire pipeline.

While promising, more work is needed in this direction before this method can be used with real data. First, the parameter space must be broadened, both for cosmology (including parameters such as neutrino masses and the Hubble constant) and astrophysics (including all parameters that can potentially affect a given observable). Second, further work is needed to make the method robust against systematic effects. Third, while in this work we have assumed that the fields can be observed without noise, this is not the case with real data. It is thus important to include the noise and instrumental effects that enter into direct observables, and retrain the network with those. Fourth, some of the fields considered in this work are not directly observable. Therefore it is important to repeat this exercise with fields that can be directly observed with cosmological surveys. Finally, given the small volume covered by our simulations, super-sample\footnote{The spatial distribution of matter, gas, galaxies, etc., is affected by large-scale perturbations not modeled in our simulations due to their small volume.} effects are not properly modeled. While naively one may expect that these effects may degrade the parameter constraints, it is not clear if this will be the case when the information is extracted from the field itself or whether the network may use that to extract additional information. Despite these challenges, we believe the potential of this approach is enormous, and can transform the manner we do cosmology.

Finally, we release all the maps from all fields used in this work in the CAMELS Multifield Dataset (CMD). We also make our codes publicly available and provide the weights of the trained networks to allow the community to reproduce our results. We formulate the problems encountered in this work as a challenge, and open it to the whole community. Instructions on how to download the data, codes, and network weights can be found at \url{https://camels-multifield-dataset.readthedocs.io}.

\matmethods{

In this section we describe the data we employ and how we generated it from the CAMELS simulations \cite{CAMELS}. We also outline the architecture of the networks used to perform parameter inference and the details of their training.

\subsection*{The CAMELS simulations}
\label{subsec:CAMELS}

We use data from the CAMELS project \citep{CAMELS}. CAMELS is a set of 4,233 numerical simulations: 2,184 state-of-the-art (magneto-)hydrodynamic simulations and 2,049 N-body simulations. The N-body simulations have been run with the \textsc{Gadget-III} \citep{Gadget} code, while the (magneto-)hydrodynamic simulations were run with two different codes: \textsc{Arepo} \citep{Arepo} and \textsc{Gizmo} \citep{Hopkins2015_Gizmo}. The simulations run with \textsc{Arepo} use the same subgrid physics as the IllustrisTNG model \citep{WeinbergerR_16a, PillepichA_16a}, while the GIZMO simulations employ the subgrid model of the SIMBA simulation \citep{Dave2019_Simba}. CAMELS contains two different simulation suites where their names represent the subgrid models employed: IllustrisTNG and SIMBA.

All simulations follow the evolution of $256^3$ dark matter particles, plus $256^3$ gas resolution elements in the case of the hydrodynamic simulations, from $z=127$ down to $z=0$, in a periodic box of comoving volume equal to $(25~h^{-1}{\rm Mpc})^3$. The values of the following cosmological parameters are kept fixed in all simulations: $\Omega_{\rm b}=0.049$, $h=0.6711$, $n_s=0.9624$, $M_\nu=0.0$ eV, $w=-1$, $\Omega_{\rm K}=0$.

All simulations vary however the value of $\Omega_{\rm m}$ and $\sigma_8$, while the hydrodynamic simulations also vary the value of four astrophysical parameters: $A_{\rm SN1}$, $A_{\rm SN2}$, $A_{\rm AGN1}$, $A_{\rm AGN2}$, where $A_{\rm SN1}$ and $A_{\rm SN2}$ control the energy/mass loading and speed of galactic winds, while $A_{\rm AGN1}$ and $A_{\rm AGN2}$ characterize the energy and jet speed of AGN feedback. We note that while the name of the astrophysical parameters is the same in  SIMBA and IllustrisTNG, e.g. $A_{\rm SN1}$, their actual implementation and physical meaning is in general different. This should be taken into account when training the networks on one simulations and testing on another; while a network trained on IllustrisTNG maps is expected to predict the value of $\Omega_{\rm m}$ from a SIMBA map, the same is not true for, e.g., $A_{\rm AGN2}$.

Each simulation suite has four different sets, LH, 1P, CV, EX, depending on how the variations of the cosmological and astrophysical parameters, as well as the initial conditions, are organized. In this work we primarily use the simulations from the IllustrisTNG LH set, which consists of 1,000 hydrodynamic simulations whose parameters are organized in a latin-hypercube. Each simulation in this set has different initial conditions. We refer the reader to \cite{CAMELS} for further details on the CAMELS project.

\begin{figure*}[t]
\centering
\includegraphics[width=1.0\linewidth]{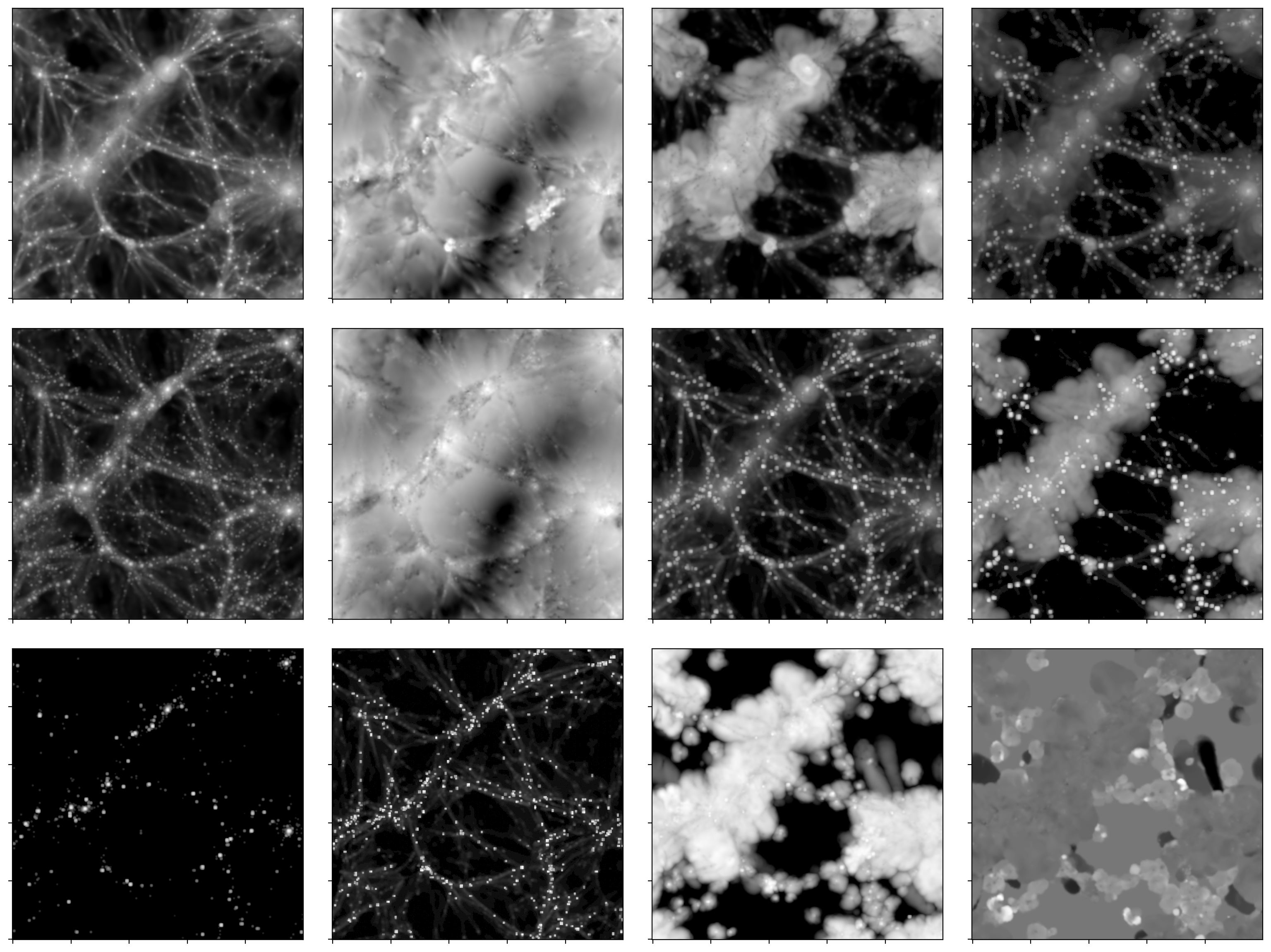}
\caption{This figure shows maps from the different fields considered. From top-left to bottom-right: gas density ($M_{\rm gas}$), gas velocity ($V_{\rm gas}$), gas temperature ($T$), gas pressure ($P$), dark matter density ($M_{\rm dm}$), dark matter velocity ($V_{\rm dm}$), electron number density ($n_e$), magnetic field ($B$), stellar mass density ($M_*$), neutral hydrogen density (${\rm HI}$), gas metallicity ($Z$), and magnesium over iron (${\rm Mg/Fe}$). In our analysis we also consider the total matter density ($M_{\rm tot}$), that receives contributions from the gas, dark matter, stellar, and black-holes components; this field is generated from both hydrodynamic simulations and gravity-only N-body simulations. The total matter density field is, visually, very similar to the dark matter density field, and we do not show it here for clarity. Each map represents a projection of a region with dimensions $25\times25\times5~(h^{-1}{\rm Mpc})^3$ along the smaller axis. We feed these images, either from a single field or from a multifield configuration, to neural networks whose goal is to perform likelihood-free inference on the value of the cosmological and astrophysical parameters. In all maps regions in black and white represent places where the value of the considered field is low and high, respectively.}
\label{fig:images}
\end{figure*}

\subsection*{Field maps}
\label{subsec:maps}

From the simulations we create 2-dimensional maps from a generic field using the following procedure. First, we select the particles within a slice of dimensions $25\times25\times5~(h^{-1}{\rm Mpc})^3$. Next, these particles and their properties are assigned to a regular 2D mesh with $256\times256$ pixels. We carry out that assignment assuming that each particle is a uniform sphere. The radii of the dark matter and gas particles is computed as the distance to their 32nd closest dark matter and gas particle, respectively. We assume that the radii of star and black hole particles are zero\footnote{We do this for simplicity since the resolution of the maps, $100~h^{-1}{\rm kpc}$, is too coarse to resolve galactic structures.}. We consider a total of 13 fields plus one multifield, as follows:

\begin{enumerate}
\item N-body: total mass density from N-body simulations. These are not contaminated by astrophysical effects.
\item $M_{\rm gas}$: gas density.
\item $M_{\rm dm}$: dark matter density.
\item $M_*$: stellar mass density.
\item $V_{\rm gas}$: gas velocity modulus.
\item $V_{\rm dm}$: dark matter velocity modulus.
\item $T$: gas temperature.
\item $Z$: gas metallicity.
\item ${\rm HI}$: neutral hydrogen density.
\item $n_e$: electron number density.
\item $P$: gas pressure.
\item $B$: magnetic field modulus.
\item ${\rm Mg/Fe}$: ratio between the magnesium and iron mass.
\item $M_{\rm tot}$: total mass (gas plus dark matter plus stars plus black holes) density. Note that this field is the same as the one in N-body, but created from hydrodynamic simulations, and therefore, contaminated by astrophysical effects.
\item ${\rm MF}$: 2D maps with 11 different channels: $M_{\rm gas}$, $V_{\rm gas}$, $M_*$, $T$, $Z$, ${\rm HI}$, $n_e$, $P$, $B$, ${\rm Mg/Fe}$, and $M_{\rm tot}$. Each channel can be interpreted as a different \textit{color} of the same image. We refer to this configuration as multifield. Note that the multifield contains all other fields with the exception of dark matter density and dark matter velocity.
\end{enumerate}
For each simulation we generate 15 maps; 5 maps in the XY plane, 5 maps in the YZ plane and 5 maps in the XZ plane. We show examples of these maps in Fig. \ref{fig:images}. While some images look rather similar (e.g. $M_{\rm gas}$, $M_{\rm dm}$, $n_e$, $P$, ${\rm HI}$), others show a very distinct morphology (e.g. ${\rm Mg/Fe}$).

The goal of this work is to train deep convolutional neural networks on these maps to perform likelihood-free inference of the value of the cosmological and astrophysical parameters. In particular, we want to quantify how accurately we can infer the value of the cosmological parameters from the different fields, which are affected by astrophysical effects in very different and distinct ways; while the effect is expected to be moderate for dark matter, for fields such as ${\rm Mg/Fe}$ and $Z$ the effect is expected to be more dramatic. Further, we want to compare our results against those that we obtain when we use fields that are not affected by astrophysics effects (using the N-body maps), and quantify how much information we can gain when using several fields at the same time (multifield).

For each field, we split the entire data set, containing 15,000 images (15 maps per simulation for 1,000 simulations) into training (900 simulations and 13,500 maps), validation (50 simulations and 750 maps), and testing (50 simulations and 750 maps). We emphasize that maps from a given simulation can only be included in the training, validation or testing set, to avoid spurious leakage from internal correlations among maps. We train the networks using data augmentation, consisting in eight unique image transformations built from rotations and flippings. 

\begin{figure*}[!t]
\centering
\includegraphics[width=1.0\linewidth]{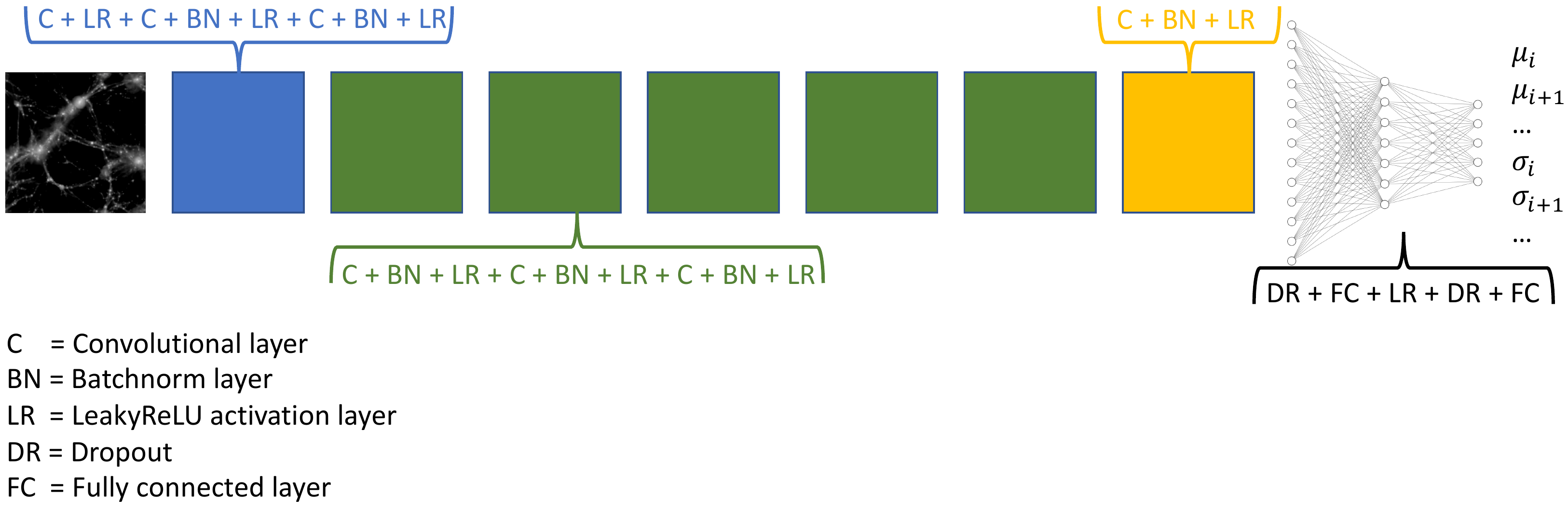}
\caption{This scheme shows the architecture of our model. The input map is pass through a set of six blocks; each block consists of a set of convolutional, batchnorm, and non-linear activation layers, as indicated. After each block, the number of channels is multiplied by two and the dimensionality of the images is reduced by a half, with the exception of the last block, where the images reduce to $1\times1$. The number of channels after the first block is a hyperparameter that we tune; the number of channels in all others convolutional layers, and the number of neurons in the fully connected layers are determined once that hyperparameter is known. After the last block, the $1\times1$ images are passed through two fully connected layers that output 12 values, the mean and variance of the posterior for each of the six parameters.}
\label{fig:architecture}
\end{figure*}

\subsection*{Neural networks}
\label{subsec:networks}

We use convolutional neural networks to find the mapping between the 2D maps and the posterior
\begin{equation}
p(\vec{\theta} | \textbf{X}) ~,   
\label{Eq:posterior}
\end{equation}
where $\vec{\theta}=(\Omega_{\rm m}, \sigma_8, A_{\rm SN1}, A_{\rm SN2}, A_{\rm AGN1}, A_{\rm AGN2})$ are the values of the parameters and $\textbf{X}$ represents a (multi)field map. We note that the N-body maps are characterized by just the two cosmological parameters. In this work we do not calculate the full posterior from Eq. \ref{Eq:posterior}, but only the mean and variance of the marginal posteriors
\begin{eqnarray}
\mu_i(\textbf{X}) &=& \int_{\theta_i} p_i(\theta_i | \textbf{X}) \theta_i d\theta_i~,\\
&& \nonumber \\
\sigma_i(\textbf{X}) &=& \int_{\theta_i} p_i(\theta_i | \textbf{X}) (\theta_i - \mu_i)^2 d\theta_i~,
\label{Eq:mean_posterior}
\end{eqnarray}
where $p_i(\theta_i|\textbf{X})$ is the marginal posterior over the parameter $i$
\begin{equation}
p_i(\theta_i|\textbf{X}) = \int_{\theta} p_i(\theta_i | \textbf{X})d\theta_1...d\theta_{i-1}d\theta_{i+1}...d\theta_n~.
\label{Eq:variance_posterior}
\end{equation}
The architecture of the model we use to carry out this task consists of a set of convolutional layers, combined with batchnorm and LeakyReLU activation layers, followed by two fully connected layers. The details of the model can be found in Fig. \ref{fig:architecture}. We choose that model because it gave us the most accurate results after trying many different models with different numbers of convolutional and fully connected layers. 

Our model takes as input a (multi)field map and returns twelve numbers, representing the mean and variance of the marginal posterior for each of the six cosmological and astrophysical parameters from Eqs. \ref{Eq:mean_posterior} and \ref{Eq:variance_posterior}. The loss function we minimize via gradient descent is 
\begin{eqnarray}
\mathcal{L}&=&\sum_{i=1}^6\log\left(\sum_{j\in{\rm batch}}\left(\theta_{i,j} - \mu_{i,j}\right)^2\right)\nonumber \\
+&&\sum_{i=1}^6\log\left(\sum_{j\in{\rm batch}}\left(\left(\theta_{i,j} - \mu_{i,j}\right)^2 - \sigma_{i,j}^2 \right)^2\right)~,
\end{eqnarray}
where the interior sum runs over all elements in the batch while the external sum runs over all six parameters. This functional form arises from the moment networks described in \cite{moment_networks}, and ensures that the twelve numbers output by the network represent the mean and variance of the marginal posterior.

We have chosen to minimize the log of each term so that all terms have the same order of magnitude. Without taking the logarithm of each term, the loss function will inevitably be dominated by the parameters that are more difficult to predict (e.g. the AGN parameters for the dark matter field). In this case, the noisy estimates of the gradients from those terms can be much larger than the ones from the parameters that are more accurately predicted, making it difficult to further improve on the accuracy of the latter. When taking the logarithm of each term, the contribution of all terms are of the same order of magnitude, facilitating the learning process.

We use the AdamW optimizer \citep{AdamW} with a cyclical learning rate policy scheduler \citep{CyclicLR}. The value of the hyperparameters (number of channels in the convolutional layers, learning rate, weight decay, and dropout rate) are tuned using the bayesian optimizer \textsc{optuna} \citep{Optuna}. For that, we run 50 trials and choose the hyperparameters values that give the lowest validation loss. Each trial trains a network with the chosen value of the hyperparameters for 500 epochs using a batch size of 128. We provide further details on the architecture of our model and on the training procedure in our companion paper \citep{CMD}.

\subsection*{Data availability} We make publicly available 1) all data used for this work, comprising tens of thousands of 2D maps for 13 different fields, 2) all codes used to train the networks, and 3) the weights of the networks themselves. All these data products can be found at \url{https://camels-multifield-dataset.readthedocs.io}. We also provide further details on how to work with this data in a companion paper \citep{CMD}.

}
\showmatmethods{}

\acknow{The training of the neural networks has been carried out using GPUs from the \textit{Rusty} and \textit{Tiger} clusters at the Flatiron Institute and Princeton University, respectively. FVN acknowledges funding from the WFIRST program through NNG26PJ30C and NNN12AA01C. DAA was supported in part by NSF grants AST-2009687 and AST-2108944. The work of SG, DAA, BW, SH, and DNS has been supported by the Simons Foundation. We have made use of the \textit{Pylians} libraries, publicly available at \url{https://pylians3.readthedocs.io}. Details about the CAMELS simulations can be found in \url{https://www.camel-simulations.org}. All maps used in this work belong to the CAMELS Multifield Dataset (CMD), publicly available in \url{https://camels-multifield-dataset.readthedocs.io}.}
\showacknow{}
\vspace{0.2cm}

\bibliography{references}

\end{document}